# Effect of Polarization Field on the Photocurrent Characteristics of GaN/AlN Multi-Quantum-Well Avalanche Photodiode

Wangping Wang,[1,a] Qian Li,[1] Xingzhao Wu,[2] Jianbin Kang,[1] Yi Luo,[2] Mo Li,[1] and Lai Wang[2]

[1]*Microsystem and Terahertz Research Center, China Academy of Engineering Physics, Chengdu 610299, People's Republic of China*

[2]*Tsinghua National Laboratory for Information Science and Technology, Department of Electronic Engineering, Tsinghua University, Beijing 100084, People's Republic of China*

Polarization is an important property of GaN/AlN multi-quantum-well (MQW) avalanche diode (MAPD) but has been ignored in recent analyses of MAPD to simplify the Monte Carlo simulation. Here, the photocurrent characteristics of GaN/AlN MAPD are investigated to understand the role of polarization field in the MQW structure. Carrier multiplication in AlN/GaN MAPD is found as a result of interfacial impact ionization not much helped from external field but instead considerably contributed by the polarization field. In addition, the movement of ionized electrons out of quantum well is proved as Fowler-Nordheim tunneling process helped by the polarization field in AlN barrier. Furthermore, the transport of ionized electrons through MQW structure is found influenced by the reverse polarization field in GaN layer, which could be suppressed by the external electric field. With all the three effects, the quick photocurrent increase of GaN/AlN MAPD is ascribed to the efficient transport of interfacial-ionized electrons step by step out of MQW at high voltages, which is in big difference from conventional APD due to avalanche breakdown.

_______________________________

[a] Electronic mail: wangwangping@mtrc.ac.cn.

[b] Electronic mail: wanglai@tsinghua.edu.cn.

Avalanche photodiode has been widely used for weak light detection.[1-3] Geiger mode avalanche photodiode is capable of single photon detection but suffered from after pulse in avalanche-quenching operation.[1-3] The avalanche-quenching circuit is not needed for avalanche photodiode with photomultiplier tube (PMT) like multiplication process, which could be realized when the electron and hole ionization rate are in large difference.[4] Greatly improved ionization rate ratio has been demonstrated in earlier AlGaAs/GaAs muti-quantum-well (MQW) avalanche photodiode (MAPD) and is attributed to the enhanced electron impact ionization by the conduction band edge discontinuity at heterointerface ($\Delta$Ec).[5-6] The $\Delta$Ec is greatly increased in GaN/AlN MQW structure and PMT-like high avalanche gain has been demonstrated for a 20 periods GaN/AlN MAPD,[7] which implies the advantage of III-nitride heterostructure for solid state PMT application.

Ensemble Monte Carlo simulation has been carried out to obtain the ionization coefficient for GaN/AlN MAPD.[7-10] However, the polarization fields in AlN and GaN layer were not considered and the interface scattering was ignored to simplify the simulation model of GaN/AlN MAPD.[7-10] It was predicted that besides the conduction band edge discontinuity, the electron ionization rate could also be enhanced by the polarization fields in III-nitride MAPD.[11] Thus the impact ionization process could not be well modeled if the polarization field in GaN/AlN MQW structure is not considered in the Monte Carlo simulation. Moreover, the directions of polarization fields in AlN and GaN layer are contrary to each other.[11] Therefore the electrons transport through the MQW structure are expected influenced by the reverse polarization field in AlN and GaN layer. To our knowledge, the carrier transport through polarized MQW structure is not well discussed in previous analyses of III-nitride MAPD.[7-11]

In this work, the photocurrent characteristic of GaN/AlN MAPD is investigated to understand the effects of polarization field on electron ionization and electron transport. At first, the GaN/AlN MAPD is theoretically investigated to get the polarization field modified energy band profile. Then a conventional APD (CAPD) is designed with similar energy band profile as MAPD but without the MQW structure. The photocurrent of MAPD is experimentally compared to that of CAPD for illustrating the effect of polarized MQW on electron ionization. Finally, the carrier transport through polarized MQW is analyzed with the rate of increase and noise feature in MAPD photocurrent.

The structure of GaN/AlN MAPD is illustrated in Fig. 1(a) and could be simply depicted as 20 AlN (10 nm)/GaN (10 nm) periodic layers inserted into the impact ionization region of p-i-p-i-n structure. The structure of homojunction CAPD is shown in Fig. 1(b) and is further described later. The MAPD and CAPD are both made of the same double-mesa device structure and device fabricating parameters for better comparison.

The role of polarization field on conduction band structure of GaN (10nm)/AlN (10 nm) MAPD is presented in Fig. 2(a). The calculation of band structure is based on self-consistent solution of Schrödinger-Poisson equations. The polarization field strength in AlN and GaN layer is obtained as 3.2 MV/cm and -3.2 MV/cm respectively, based on the software package of Crosslight APSYS with Coulomb screening ratio set as 0.5. For the MAPD intentionally ignoring the polarization field, the energy band diagram is shown as tilted profile dropped down by the built-in electric field and external electric field. In contrast, for the MAPD considering the polarization field, uplifted energy band profile is found for the MAPD structure, with nearly-flat energy band profile in the light-absorption layer. Note that the polarization field in GaN layer is in reverse direction contrary to the external electric field and built-in electric field of MAPD structure. As shown in Fig. 2(b), the electric field strength in the light-absorption layer is changed from positive to negative by the internal polarization field, which explains the nearly-flat band profile in the absorption layer. In the latter part of the paper, a comparative CAPD is designed with similar band profile as polarized MAPD but without the MQW structure to fully illustrate the effect of polarized MQW. In consideration of the nearly-flat band profile in the MAPD absorption layer, the CAPD should be designed as p-i-n structure but not as the same p-i-p-i-n structure of MAPD. The comparison could be made since the nearly-flat band profile in absorption layer of MAPD is found very stable regardless of the bias voltage.

In Fig. 3(a), the photocurrent of GaN/AlN MAPD is compared with that of GaN CAPD to obtain the role of polarization field on ionization process. The CAPD is designed with intrinsic layer thickness as 20 GaN periodic layers, or half of the MQW thickness in MAPD. At same bias voltage, the external field strength in MAPD is about half of that in CAPD, but the photocurrent of MAPD is orders larger than that of CAPD. This is an indication of the structure-induced carrier multiplication in MAPD. Furthermore, increase of more than one order is observed in the photocurrent of MAPD applied to 10 V, while negligible increase is shown in the photocurrent of CAPD applied to 60 V. Hence the carrier multiplication could be started without much help from the external electric field on MAPD. In GaN/AlN MAPD, aside from external field, the electrons are also heated by the 3.2 MV/cm polarization field in AlN layer, 2.0 eV Δ Ec at GaN/AlN interface, less than 0.1 MV/cm build-in electric field of p-i-p-i-n doping, but are cooled down by the -3.2 MV/cm polarization field in GaN layer. The build-in electric field is negligible compare to the large polarization field in GaN/AlN MAPD. Therefore, at low voltages the enhanced impact ionization in MAPD is mainly resulted from the combined action of polarization field in AlN layer and $\triangle$ Ec at GaN/AlN interface. Accordingly, the impact ionization process in MAPD could be described as interfacial impact ionization at GaN/AlN interface. Note that the energy for electrons to trigger ionization events is about 3.7 eV in GaN layer.[9] For electrons from AlN to GaN at 60 V high voltage, the gained electron energy is 1.5 eV from 1.5 MV/cm external field and 2.0 eV from the GaN/AlN $\triangle$ Ec, which combined together is still smaller than the 3.7 eV ionization energy. Hence unlike CAPD in which the impact ionization is fully due to external electric field, the interfacial impact ionization in MAPD is considerably contributed by the polarization field. There is no external field induced impact ionization in GaN well layer of MAPD, for the quantum well layer is designed very thin to allow just one impact ionization triggered here.

The polarization field is not only beneficial to the electron ionization in MAPD, but is also helpful to the movement of electrons out of quantum well. As shown in Fig. 4, electron trapping at GaN/AlN interface is expected due to the nearly-flat band profile in absorption layer and tilted-up band profile in quantum well layer. Considering that the band profile of AlN layer is greatly dropped down by the polarization field, the electrons may transport through the AlN barrier in Fowler-Nordheim (FN) tunneling process. The FN tunneling current is proportional to $\varepsilon^2 \exp(-A/\varepsilon)$, where $\varepsilon$ is electric field on tunneling barrier.[12] Accordingly, the FN tunneling probability is greatly enhanced by the 3.2 MV/cm polarization field in AlN barrier. Interestingly, there is evidence for the proposed tunneling transport in the photocurrent characteristic of Fig. 3(b). Negative differential conductance (NDC) is clearly observed in MAPD photocurrent curve, with peak current about three times of valley current and peak voltage around 11 V, and is confirmed with repeated measurements of many MAPD chips. For the photocurrent of MAPD, the impact ionization is only triggered by the photoelectrons injected into MQW structure. The photoelectron injection efficiency could be greatly enhanced when photoelectrons resonant tunnel through the AlN barriers as indicated by the arrow in Fig. 4. Confined states in each GaN quantum well is obtained by self-consistently solving Schrödinger equation for the whole MQW structure. As shown in the inset figures of Fig. 4, for the 1st GaN quantum well in equilibrium condition, discrete energy levels are found above conduction band edge of light-absorption layer ($Ec_{photo}$). These discrete levels drop down and in alignment with $Ec_{photo}$ around 10V bias, which is close to the observed NDC peak voltage. Hence, the NDC feature is attributed to the triangular double barrier resonant tunneling of photoelectrons from absorption layer into MQW structure.

The resonant tunneling is possible because the confined states of GaN quantum well are greatly lifted up by the polarization field.

It is expected that the ionized electrons in each GaN well layer is step by step multiplied at following GaN/AlN interface. The transport of ionized electrons is influenced by the reverse polarization field in GaN layer and is investigated with different rates of increase of the MAPD photocurrent in Fig. 3(b). As shown in the linear scale photocurrent curve, there are two rates of increase of the MAPD photocurrent: slow increase labeled as rate 1 and quick increase labeled as rate 2, with transition voltage around 33 V between rates 1, 2. Unlike the sharp increase of CAPD photocurrent due to avalanche breakdown, the quick increase of MAPD photocurrent is observed to be saturated above 65 V, hence rules out the possibility of avalanche breakdown in MAPD. Moreover, it is observed that feature of large noise is appeared in rate 1 part of MAPD photocurrent curve and is gradually reduced in rate 2 part of photocurrent with increasing bias voltage. The noise feature is confirmed with repeated measurements in many chips of GaN/AlN MAPD. In contrast, no such noise feature is observed in the photocurrent curve of CAPD. Thus the noise characteristic of MAPD is correlated with the polarized MQW structure and is not related with device processing parameters or measurement equipment.

In view of the noise feature, the different rate of increase of the MAPD photocurrent is explained with arrow-indicated electron steam through the polarized MQW structure in Fig.5. There are two kinds of electrons inside each quantum well layer: interfacial-ionized electrons created in current quantum well and drifting electrons from previous quantum well to current quantum well. As indicated by arrow 1, the interfacial-ionized electrons are partly trapped and scattered by the reverse polarization field in quantum well layer while the drifting electrons are scattered at previous and current AlN barriers. Note that the AlN barriers are uplifted in the band profile also due to the reverse polarization field in GaN layer. Hence the photocurrent is observed both of noisy feature and slow rate of increase below 33 V, because of the large carrier trapping and interface scattering for arrow 1 indicated electron stream. At high voltages, not only the reverse polarization field in GaN well layer is partly offset by the external field, but also the FN tunneling through AlN barrier is enhanced. Furthermore, as indicated by arrow 2, the drifting electrons are easily heated over AlN barrier at high voltage, thereby minimizing the influence of MQW structure on carrier transport. Thus quick increase of photocurrent is observed with gradually reduced noise above 33 V, thanks to the greatly reduced carrier trapping and interface scattering in MQW structure. As indicated by the dash arrow in Fig. 5, the AlN barrier is dropped just below the previous quantum well at 33 V, which sets the transition voltage for arrows 1, 2 indicated electron stream and rates 1, 2 indicated photocurrent.

From the above analysis, although the reverse polarization field in GaN layer enhances the carrier trapping and interface scattering in MQW structure, the multiplication gain of GaN/AlN MAPD is influenced negligibly by the polarization field in GaN layer. In consideration of the photocurrent characteristics in Fig. 3(a), the best detection voltage of GaN/AlN MAPD is set around 60 V, when the reverse polarization field is greatly suppressed and the interfacial-ionized electrons are efficiently transported step by step out of MQW to collector contact. Moreover, the polarization field in GaN layer is found inverse proportional to its layer thickness.[11] Accordingly, the GaN/AlN MAPD could be further optimized to reduce the reverse polarization field by adopting thick GaN well layer. The carrier trapping and interface scattering noise of MAPD are also reduced with this optimization.

In summary, the photocurrent characteristic of GaN/AlN MAPD has been theoretically and experimentally investigated. The polarization field in AlN layer is not only confirmed of large contribution to the enhanced ionization rate in MAPD but is also helpful to the movement of electrons out of quantum well layer. The polarization field in GaN layer has influence on the transport of ionized electrons through MQW structure but could be suppressed by the external electric field. Although the avalanche gain is currently smaller than CAPD, the GaN/AlN MAPD has unique advantages over CAPD. Besides the demonstrated PMT-like detection mode of MAPD without avalanche-quenching circuits, the GaN/AlN MAPD is expected of better responsivity spectral cutoff over CAPD. Due to the reverse polarization field in GaN layer, as shown in Fig. 2(b), the light absorption layer in GaN/AlN MAPD is found with electric field strength lower than $5 \times 10^4$ V/cm, which is orders smaller than the typical $10^6$ V/cm field strength in the absorption layer of CAPD. Therefore, the Franz-keldysh effect is minimized in the absorption layer of III-nitride MAPD, which should improve the responsivity spectral performance of III-nitride MAPD at high detection bias.

## ACKNOWLEDGMENTS

This work is supported by the National Natural Science Foundation of China (Grant Nos. 61574082).

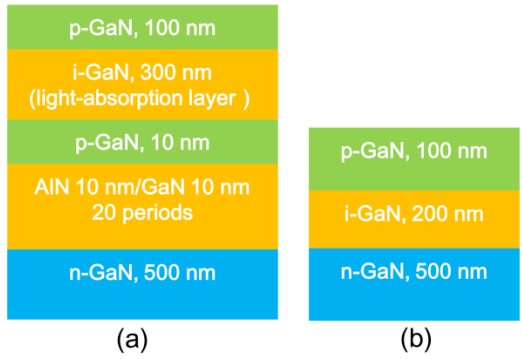

FIG.1 Material structure of (a) GaN/AlN MAPD and (b) GaN CAPD.

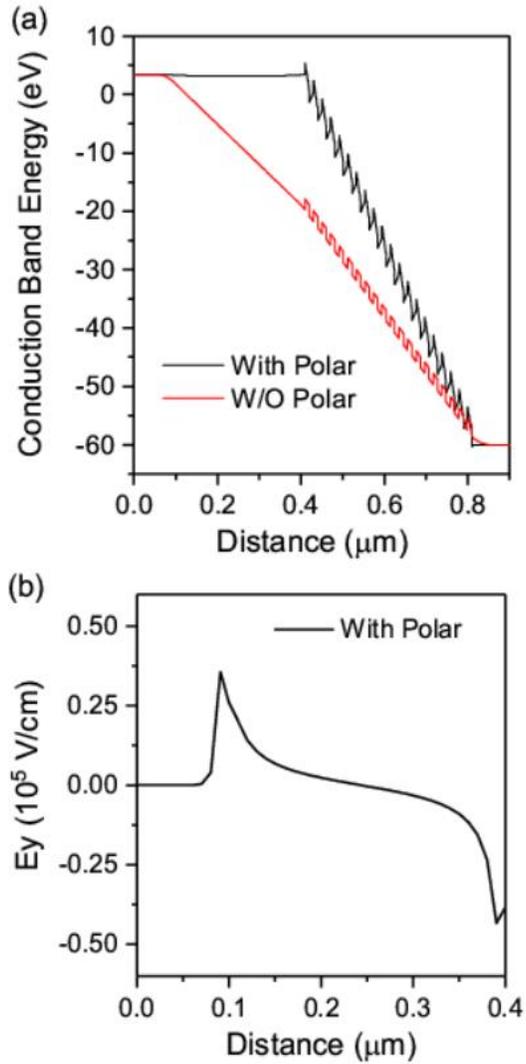

FIG.2 (a) Conduction band profile of GaN/AlN MAPD at 60 V voltage with or without considering polarization field. (b) Electric field distribution in 60 V biased MAPD considering polarization field.

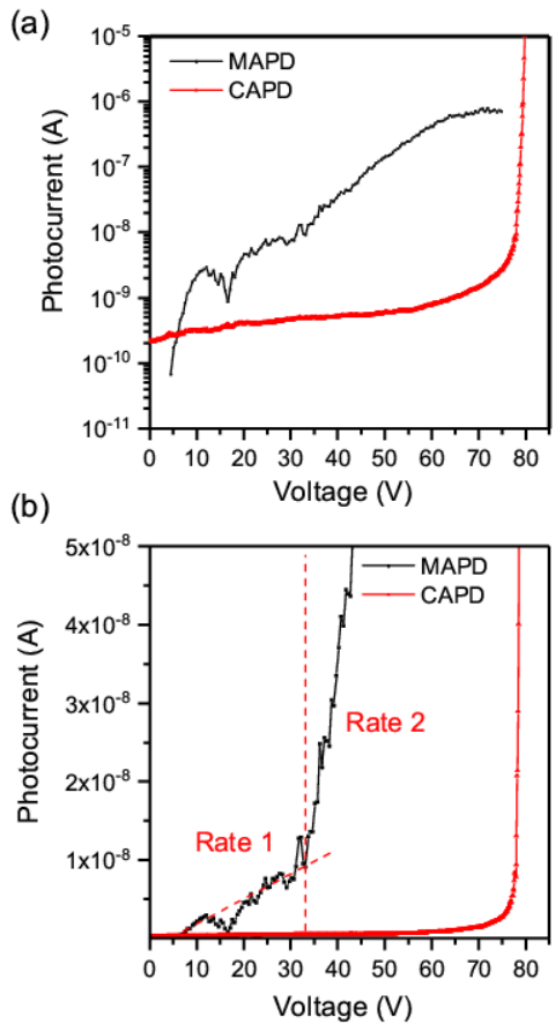

FIG.3 Photocurrent characteristics of MAPD and CAPD (a) in log-scale (b) in linear scale. The photocurrent is obtained by subtracting dark current from light current.

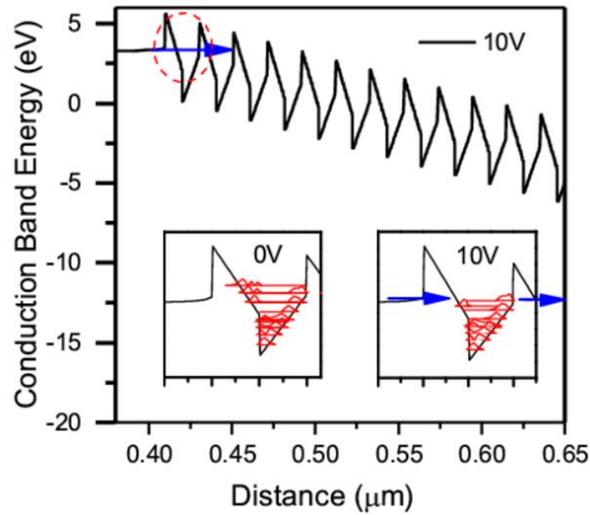

FIG.4 Calculated conduction band profile for MAPD in 10 V voltage bias. Inset figures show the confined energy levels of the GaN quantum well (circle part of main figure) in equilibrium condition and in energy alignment condition.

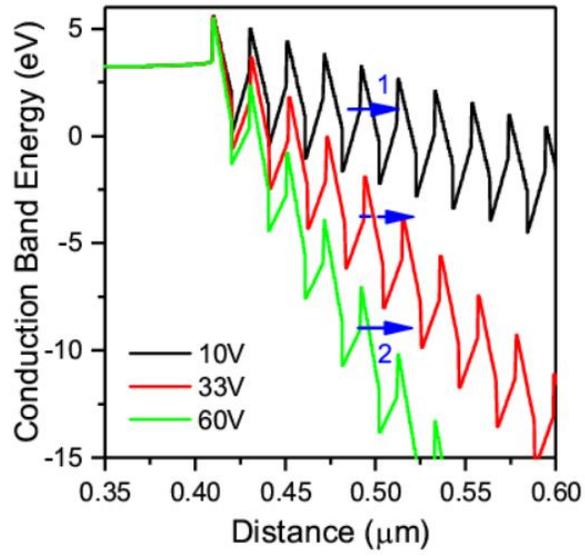

FIG.5 Calculated conduction band profile for the GaN/AlN MAPD in 10 V, 33 V, and 60 V voltage bias. The arrows indicate electron stream through MQW structure.